\documentclass[10pt,conference]{IEEEtran}
\usepackage{cite}
\usepackage{amsmath,amssymb,amsfonts}
\usepackage{algorithmic}
\usepackage{graphicx, float}
\usepackage{textcomp}
\usepackage{xcolor}
\def\BibTeX{{\rm B\kern-.05em{\sc i\kern-.025em b}\kern-.08em
    T\kern-.1667em\lower.7ex\hbox{E}\kern-.125emX}}

\usepackage[breaklinks=true, colorlinks=true, linkcolor={blue!90!black}, citecolor={blue!90!black}, urlcolor={blue!90!black}]{hyperref}
\usepackage{orcidlink}
\usepackage{comment}
\usepackage{physics}
\usepackage{qcircuit}
\usepackage{mathtools}
\usepackage{caption}
\usepackage{subcaption}
\usepackage{cleveref}
\Crefname{figure}{Fig.}{Figs.}
\usepackage{lipsum}
\usepackage{widetext}
\newcommand{\Gate}[1]{\textsc{#1}}
\newcommand{\hgate}{\Gate{h}}

\newcommand{\cnotgate}{\Gate{cnot}}

\newcommand{\swapgate}{\Gate{swap}}

\newcommand{\SB}{S_{\text{B}}}

\begin{document}
\bstctlcite{IEEEexample:BSTcontrol} %

\title{Characterizing Error Mitigation by\\Symmetry Verification in QAOA}

\author{\IEEEauthorblockN{Ashish Kakkar\orcidlink{0000-0003-1268-5417}\IEEEauthorrefmark{1},  Jeffrey Larson\orcidlink{0000-0001-9924-2082}\IEEEauthorrefmark{2}, Alexey Galda\orcidlink{0000-0001-9666-1337}\IEEEauthorrefmark{3}, and Ruslan Shaydulin\orcidlink{0000-0002-8657-2848}\IEEEauthorrefmark{2}}
\IEEEauthorblockA{\IEEEauthorrefmark{1}Dept. of Physics and Astronomy, University of Kentucky, Lexington, KY, USA}
\IEEEauthorblockA{\IEEEauthorrefmark{2}Mathematics and Computer Science, Argonne National Laboratory, Lemont, IL, USA}
\IEEEauthorblockA{\IEEEauthorrefmark{3} Menten AI, Inc., San Francisco, CA, USA}
Email: \IEEEauthorrefmark{1}ashish.kakkar@uky.edu, \IEEEauthorrefmark{2}\{jmlarson,rshaydulin\}@anl.gov, \IEEEauthorrefmark{3}alexey.galda@menten.ai}

\maketitle

\begin{abstract}
Hardware errors are a major obstacle to demonstrating quantum advantage with the quantum approximate optimization algorithm (QAOA). Recently, symmetry verification has been proposed and empirically demonstrated to boost the quantum state fidelity, the expected solution quality, and the success probability of QAOA on a superconducting quantum processor. Symmetry verification uses parity checks that leverage the symmetries of the objective function to be optimized. We develop a theoretical framework for analyzing this approach under local noise and derive explicit formulas for fidelity improvements on problems with global $\mathbb{Z}_2$ symmetry. We numerically investigate the symmetry verification on the MaxCut problem and identify the error regimes in which this approach improves the QAOA objective. We observe that these regimes correspond to the error rates present in near-term hardware. We further demonstrate the efficacy of symmetry verification on an IonQ trapped ion quantum processor where an improvement in the QAOA objective of up to 19.2\% is observed.
\end{abstract}


\section{Introduction}
Quantum computers have the potential to demonstrate a computational advantage in the near term over state-of-the-art classical algorithms~\cite{PreskillNISQ2018}. In the recent years, the size and the quality of quantum processors have  increased~\cite{ibm_2020} and their computational power has become competitive with that of classical supercomputers on certain tasks~\cite{Arute2019,Wu2021}. However, near-term quantum devices are still characterized by high error rates that impose limits on their computational power. Although in the long term quantum error correction provides a solution to this problem~\cite{preskill1997faulttolerant}, the currently available devices are far from satisfying the preconditions for error correction. This situation motivates the development of error mitigation techniques, which provide some degree of robustness to errors while avoiding the high overhead of full error correction.

The quantum approximate optimization algorithm (QAOA) is one of the most promising near-term algorithms with the possibility of showing a quantum advantage \cite{farhi2014quantum,2021Wurtz,Wurtz2021}. QAOA approximately solves a combinatorial optimization problem by preparing a parameterized quantum state. The parameters are chosen such that upon the measurement of the state, an approximate solution to the target problem is retrieved with high probability. These parameters are typically obtained by using some classical numerical optimization method. A major motivation for the study of QAOA is the existence of worst-case performance bounds that are competitive with classical local search approximation algorithms \cite{farhi2014quantum,2021FarhiSKmodelhighgirth}. 

The presence of noise on current quantum devices has a drastic effect on the applicability of QAOA. Indeed, recent results demonstrate that under realistic noise assumptions and in the absence of error mitigation, QAOA cannot outperform even a simple classical competitor on modestly sized problems~\cite{StilckFrana2021}.  Moreover, recent studies \cite{xue2019effects,Marshall_2020} show 
the effects of noise on QAOA parameters and the objective landscape: under local noise, the fidelity of the noisy state decays exponentially with the number of qubits, $N$, on the order of $\left( 1-p \right)^{N}$, where $p$ characterizes the error rate. %
Wang et al.~\cite{2021NIBP} have noted that because of the effects of local noise, gradients of the QAOA objective with respect to QAOA parameters vanish while the general landscape does not change much. The vanishing of the gradients makes it harder to optimize the QAOA parameters. Therefore, error mitigation is a prerequisite for realizing the potential of QAOA.

Many error mitigation strategies have been proposed for near-term quantum devices~\cite{errormitigationreview2021}. Improving the estimate of the expectation of some observable is the goal of several recently proposed techniques. These include zero-noise extrapolation~\cite{2017emshortdepth,digitalzne2020}, probabilistic error cancellation \cite{PEClatest}, and virtual distillation \cite{2022shadowdistillation}. Quantum subspace expansion~\cite{subspaceexpansion2020} uses classical postprocessing to remove the overhead of explicitly performing parity checks on the hardware and decoding the errors from the measurement outcomes. 

While just improving the estimate of the expectations of the observables is sufficient in some cases, often an improvement in the fidelity of the quantum state is also necessary. This is the problem we address in this work. The approach we consider is symmetry verification (SV), which was initially proposed for the variational quantum eigensolver~\cite{Bonet-Monroig,Sagastizabal2019}. Symmetry verification mitigates hardware errors by performing parity checks based on the symmetries of the problem Hamiltonian and postselecting based on the measurement outcomes. For example, if the problem Hamiltonian $H$ commutes with a given Pauli string $P$, we say that $P$ is the symmetry of $H$. When studying $H$ on a noisy quantum computer, we can restrict the quantum state evolution to one eigenspace of $P$. Then the value (``parity'') of $P$ can be measured and the state postselected on the measurement outcome corresponding to the that eigenspace of $P$.

It has been shown that symmetries of the classical objective function to be optimized are preserved by the QAOA ansatz~\cite{Shaydulin2012}. These symmetries can therefore be verified by using parity checks, with the postselection enforcing the symmetry. This approach was applied to QAOA for the MaxCut problem and was demonstrated to improve the quantum state fidelity on a cloud-based superconducting IBM quantum processor~\cite{shaydulin2021error}. 

In this work we provide analytical and numerical evidence for the power of symmetry verification to mitigate the errors in QAOA evolution when performed on a noisy quantum device. We begin by developing a theoretical framework for analyzing the fidelity improvements from symmetry verification under local noise. We use this framework to provide explicit formulas for the fidelity improvements in the limit of noise-free parity checks. We then analyze symmetry verification numerically under a more realistic noisy checks assumption, and we identify a regime in which the symmetry verification leads to QAOA objective improvements. This regime correspond to the error rates expected on near-term hardware. We verify the expected efficacy of symmetry verification on real quantum devices by performing experiments on IonQ trapped-ion processors and observing QAOA objective improvements of up to 19.2\%.

The rest of the paper is organized as follows. \Cref{sec:background} introduces QAOA, problem symmetries, symmetry verification, and our model for device noise. \Cref{sec:analytics} considers the case of noise-free parity checks and local noise and presents analytical results on the fidelity improvement from symmetry verification. \Cref{sec:higher order permutations} outlines the procedure for verifying permutation symmetries inherited from problem instances. \Cref{sec:numerics} demonstrates QAOA objective improvements from symmetry verification numerically under a specific error model, with noise applied to both the QAOA circuit and the parity checks. \Cref{sec:ionQ} presents the results obtained on IonQ trapped-ion quantum processors.

\section{Background}
\label{sec:background}

Consider the objective function $c(x)$. We study the problem of finding $ x \in \{ 0, 1 \}^{N}$ that maximizes $c$: 
\begin{equation}
    \max_{x \in \{ 0, 1 \}^{N}}c(x).
\end{equation}
Our results are general and apply to a wide range of objective functions with symmetries. The example problem we study numerically is MaxCut: Given a  graph $G$ with nodes $V$ and edges $E$, find a partition of the nodes into two sets so that the number of edges between the sets is maximized. MaxCut on general graphs is APX-complete; that is, there does not exist a polynomial-time classical approximation algorithm for MaxCut unless $\text{P}=\text{NP}$~\cite{papadimitriou1991optimization}. 

The classical objective function $c(x)$ on $N$ bits is represented by a diagonal operator (Hamiltonian) in the computational basis as
\begin{equation}\label{cost hamiltonian def}
H= \sum_{x \in \{ 0, 1 \}^{N}} c(x)\ket{x}\bra{x}.
\end{equation}
For MaxCut problem instances, this is
\begin{equation}\label{MaxCut def}
H=\frac{1}{2}\sum_{(i,j) \in E} \left( I-Z_iZ_j\right),
\end{equation}
where $Z_i$ denotes the Pauli $Z$ operator on the $i$th qubit and the identity operator on the rest of the qubits.

\subsection{The quantum approximate optimization algorithm}

QAOA approximately solves optimization problems by preparing a parameterized quantum state by acting on the initial state  $\ket{+}^{\otimes N}$ with operators $U_B(\beta_i) = e^{- i \beta_i B}$ and $U_H(\gamma_i) = e^{- i \gamma_i H}$:
\begin{equation}
\ket{\vec{\beta},\vec{\gamma}}_d = U_B(\beta_d) U_H(\gamma_d)...U_B(\beta_1) U_H(\gamma_1) \ket{+}^N.
\end{equation}
The initial state $\ket{+}^{\otimes N}$ is the ground state of the operator $B= \sum_{j=1}^n X_j$. Typically, a classical numerical optimization method is used to identify parameters $\vec{\beta},\vec{\gamma}$ that minimize the expectation value 
\begin{equation}
\label{eq:obj}
 \bra{\vec{\beta},\vec{\gamma}}_d H \ket{\vec{\beta},\vec{\gamma}}_d.
\end{equation}
We refer to the value of \eqref{eq:obj} as the ``QAOA objective'' or simply the ``objective.''
The approximation ratio for QAOA with a given set of $\vec{\beta},\vec{\gamma}$ at depth $d$ is defined as $ \bra{\vec{\beta},\vec{\gamma}}_d H \ket{\vec{\beta},\vec{\gamma}}_d/C_{\text{max}}$, where $C_{\text{max}}=\max_{x \in \{ 0, 1 \}^{N}}c(x)$.
\subsection{Symmetries in QAOA}

A symmetry of the objective function $c(x)$ is a permutation acting on binary strings $s:x\longrightarrow s(x)$ that leaves the objective unchanged, that is,  $c(s(x)) = c(x)$ $\forall x\in\{0,1\}^N$. For MaxCut on a graph $G=(V,E)$, we consider two classes of symmetries. The first is the $\mathbb{Z}_2$ symmetry, which is present for any graph $G$ and acts as a bit flip on $x$. The second class of symmetries we consider is problem instance specific. These symmetries are permutations of the indices of the bit string $x$, that is, $s\in S_N$. If a permutation $s$ satisfies $(s(u),s(v)) \in E$ when $(u,v) \in E$, $s$ is a symmetry of the MaxCut objective for $G$.

A transformation $s$ that leaves the objective function invariant is represented on the $N$-qubit Hilbert space by the operator
\begin{equation}
S = \ket{s(x_1) s(x_2) \ldots s(x_N)}\bra{x_1 x_2 \ldots x_N}.
\end{equation}
In particular, the bit-flip symmetry operator is represented by
\begin{equation}
\SB = \prod_{i} X_i \quad \forall i \in V.
\end{equation},
and the permutation on two qubits $(i,j)$ (i.e., the $\swapgate$ gate) is represented by 
\begin{equation}
S_{i,j}= \frac{1}{2} \left( I_i \otimes I_j + X_i \otimes X_j + Y_i \otimes Y_j + Z_i \otimes Z_j  \right).
\end{equation}
QAOA preserves the symmetries of the objective function~\cite{Shaydulin2012}.
Concretely, Theorem 1 in \cite{Shaydulin2012} states that \emph{the QAOA ansatz $\ket{\vec{\beta},\vec{\gamma}}$ is stabilized by a symmetry operator $S$ if $[S,B]=0$, $[S,H]=0$, and $S \ket{+}^{\otimes N}= \ket{+}^N$}. 
We will discuss arbitrary permutation symmetries in \Cref{sec:higher order permutations}; for now we focus on $\SB$ and $S_{i,j}$.  The initial state $\ket{+}^N$ is stabilized by $\SB$ and arbitrary permutations, including $S_{i,j}$. Moreover, $[B, \SB]=0=[B, S_{i,j}]$ trivially. 
As a result, for any $G$, the MaxCut QAOA ansatz is stabilized by $\SB$. If $G$ is invariant under the permutation of nodes $(i,j)$, then the corresponding ansatz is stabilized by $S_{i,j}$.

\subsection{Symmetry verification}

We now briefly review error mitigation by symmetry verification \cite{Bonet-Monroig}. If the initial state is in the $+1$ eigenspace of a symmetry $S$, then in the absence of errors it must stay in this eigenspace during circuit evolution. If errors occur during the evolution, however, the state may not remain in the $+1$ eigenspace of $S$. Projective measurements of $S$ can then be used to detect errors by checking whether the state is in the correct eigenspace after the noisy circuit evolution. 
To construct the set of projective measurement operators for the symmetry $S$, we consider its spectral decomposition. Because both $\SB$ and  $S_{i,j}$  satisfy $S^2=I$, we can write
\begin{equation}
S = \sum_{s=\pm 1} s P_s,
\end{equation}
where  $P_{s} = \frac{1}{2} \left( I + s S \right)   $ are projectors onto the eigenspace corresponding to eigenvalues $s=\pm 1$.

\begin{figure}
\[
\Qcircuit @C=1em @R=1em{
        \lstick{\ket{+}}  & \multigate{3}{\text{QAOA}} & \qw & \multigate{3}{S}    & \qw   & \qw   & \\
        \lstick{\ket{+}} & \ghost{\text{QAOA}}        & \qw & \ghost{\text{\textbf{S}}}  & \qw    & \qw \\
       \lstick{\ket{+}}  & \ghost{\text{QAOA}}        & \qw        & \ghost{\text{\textbf{S}}}   & \qw    & \qw \\
        \lstick{\ket{+}} & \ghost{\text{QAOA}}        & \qw        & \ghost{\text{\textbf{S}}}   & \qw    & \qw\\
        \lstick{\ket{0}} & \qw        & \gate{\hgate} & \ctrl{-1}  &  \gate{\hgate} & \meter & \cw
}
\]
\caption{Verification of symmetry $S$ on the QAOA state. Postselecting on the ancilla measurement outcome $0$ projects the state $\ket{\psi}$ to $P_{+1}\ket{\psi}$. \label{fig:generic_SV_circ}}
\end{figure}
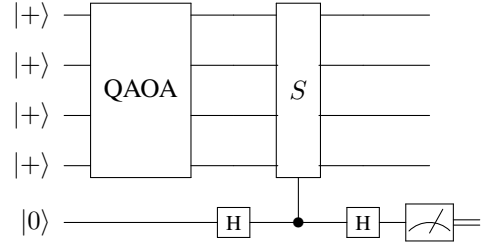

Let $\ket{\psi}\bra{\psi}$ denote the (hypothetical) noise-free state prepared by the circuit, and let $\rho_{\text{noisy}}$ denote the state with errors. A projective measurement of $S$ is performed, and only measurements corresponding to the $+1$ eigenvalue are kept. The corresponding circuit is shown in \Cref{fig:generic_SV_circ}. This is equivalent to the projector $P_{+1}$ acting on the prepared state. This process has the following important effect: the postselected state $\rho_{\text{SV}} = \frac{\left(P_{+1} \rho_{\text{noisy}} P_{+1}\right)}{\Tr \left(P_{+1} \rho_{\text{noisy}}  \right)}$  has a greater overlap with the error-free state $\ket{\psi}\bra{\psi}$. This overlap is captured by the fidelity of the state after SV, denoted by $F_{\text{SV}}$:
\begin{equation}
F_{\text{SV}} = \Tr \left(\rho_{\text{SV}} \ket{\psi}\bra{\psi} \right) =   \frac{\Tr \left(\rho_{\text{noisy}} \ket{\psi}\bra{\psi} \right)}{\Tr\left(  P_{+1} \rho_{\text{noisy}} \right)}.   
\end{equation}
Noting that the fidelity of the noisy state without error mitigation is $ F_\text{noisy} =\Tr \left(\rho_{\text{noisy}} \ket{\psi}\bra{\psi} \right)$, we can write the ratio 
\begin{equation}\label{ratio fidelity}
\frac{F_{\text{SV}}}{F_\text{noisy}}= \frac{1}{\Tr \left( P_{+1} \rho_{\text{noisy}}\right)}.
\end{equation}
This ratio is a measure of the improvement in the fidelity of the postselected state due to symmetry verification. We denote $F_{\text{SV}}$ by $F_{\SB}$ when the symmetry being verified is $\SB$ and by $F_{S_{i,j}}$ if the symmetry is $S_{i,j}$.

\subsection{Modeling the errors in QAOA}
For our theoretical analysis we model the noise occurring in the QAOA evolution as local (single-qubit) error channels. Specifically, we follow a commonly used error model that applies a layer of a single-qubit noise channel after each pair of QAOA operators (i.e., a ``QAOA layer'') \cite{Marshall_2020,xue2019effects,2021NIBP}. We study the physically relevant depolarizing and dephasing noise channels. Let $\rho = \ket{\psi}\bra{\psi}$ denote the (noiseless) pure state prepared by a single layer of QAOA. %
Our model assumes this state is acted upon by the noise channel $\mathcal{E}_i$ on the $i^{th}$ qubit. 

The single-qubit depolarizing noise channel is defined by the set of Kraus operators $E_a \in \{ \sqrt{(1-p)}I, \sqrt{\frac{p}{3}} X,\sqrt{\frac{p}{3}} Y,\sqrt{\frac{p}{3}} Z  \}$ and acts on $\rho$ as 
\begin{equation} \label{noisy channel 1 qubit}
\mathcal{E}_i(\rho)=  \left( 1-  p  \right) I \rho I + \left(\frac{p}{3}  X_i \rho X_i +\frac{p}{3}  Y_i \rho Y_i +\frac{p}{3} Z_i \rho Z_i \right).
\end{equation}
We additionally consider the single-qubit dephasing noise channel acting as
\begin{equation} \label{noisy channel 1 qubit dephasing}
\mathcal{E}_i(\rho)=  \left( 1-  p  \right) I \rho I + \left(p Z_i \rho Z_i \right).
\end{equation}
Here  $p \leq 1 $, and the normalizations are chosen to ensure that the map is trace preserving, namely, $ \sum_a E_a^\dagger E_a = I$. 
\par
\begin{figure*}
\[
\Qcircuit @C=1em @R=1em {
        \lstick{\ket{+}} & \multigate{3}{e^{-i\gamma_1 H}}  & \multigate{3}{e^{-i\beta_1 B}} & \gate{\mathcal{E}} &  \multigate{3}{e^{-i\gamma_2 H}} & \multigate{3}{e^{-i\beta_2 B}} & \gate{\mathcal{E}} & \multigate{3}{S} & \qw \\
        \lstick{\ket{+}} & \ghost{e^{-i\gamma_1 H}}         & \ghost{e^{-i\beta_1 B}}        & \gate{\mathcal{E}} & \ghost{e^{-i\gamma_1 H}}         & \ghost{e^{-i\beta_1 B}}        & \gate{\mathcal{E}} & \ghost{A}                        & \qw \\
        \lstick{\ket{+}} & \ghost{e^{-i\gamma_1 H}}         & \ghost{e^{-i\beta_1 B}}        & \gate{\mathcal{E}} & \ghost{e^{-i\gamma_1 H}}         & \ghost{e^{-i\beta_1 B}}        & \gate{\mathcal{E}} & \ghost{A}                        & \qw \\
        \lstick{\ket{+}} & \ghost{e^{-i\gamma_1 H}}         & \ghost{e^{-i\beta_1 B}}        & \gate{\mathcal{E}} & \ghost{e^{-i\gamma_1 H}}         & \ghost{e^{-i\beta_1 B}}        & \gate{\mathcal{E}} & \ghost{A}                        & \qw \\
        \lstick{\ket{0}} & \qw                            & \qw                           & \qw                & \qw                            & \qw                           & \gate{\hgate}      & \ctrl{-1}                        & \gate{\hgate} & \meter  
}
\]
\caption{Error model of local noise acting between each layer of QAOA followed by verification of symmetry $S$. $\mathcal{E}$ is a noise channel that acts on each qubit locally. \label{fig:circuit2}}
\end{figure*}
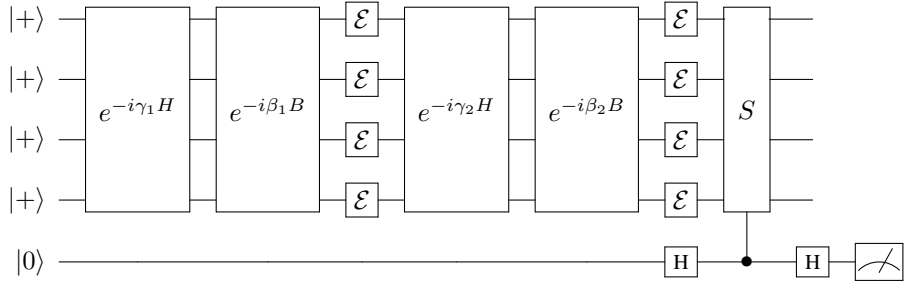

{}
These single-qubit noise channels act on all qubits, with the state after a single layer of QAOA given by
\begin{equation} \label{full error model 1 layer}
\rho_1 = \mathcal{E} \big( \rho \big)=\prod_{i=1}^N \mathcal{E}_{i} \big( \rho \big),
\end{equation}
where each $\mathcal{E}_{i}$ acts independently. Note that only the noise operator $\mathcal{E}$ factorizes over qubits; the state $\rho$ is a general $N$-qubit state. The noise channel is then applied repeatedly after each QAOA layer. We can generalize to $d$ layers of QAOA by writing the action of each QAOA layer (indexed by $t$) as $\mathcal{U}^{(t)} \left(\beta_{t},\gamma_{t}\right)$ and the noise channel between each layer as $\mathcal{E}=\prod_{i=1}^N \mathcal{E}_{i}$. We arrive at the following density matrix for the noisy state at depth $d$: 
\begin{equation} \label{full noise model}
\rho_d =\mathcal{E}  \circ  \mathcal{U}^{(d)} \circ  \cdots \circ  \mathcal{E}  \circ  \mathcal{U}^{(1)} \circ \big( \ket{+}^{\otimes N}\bra{+}^{\otimes N} \big).
\end{equation}
A high-level overview of the noise model is given in \Cref{fig:circuit2} with SV performed after the final layer.

\section{Mitigating the effects of local noise on QAOA}\label{sec:analytics}

Recent results~\cite{xue2019effects,Marshall_2020,StilckFrana2021,shaydulin2021error}  show that using error mitigation strategies to minimize the effects of local noise is necessary in order to achieve quantum advantage with QAOA. Here we  demonstrate that error mitigation via symmetry verification improves the fidelity of the QAOA state under local noise. We provide explicit formulas for the relationship between the fidelity improvement, probability of local noise, and  circuit depth and size. Our formulas rely on the assumption that no noise is present during the decoding procedure (i.e., the parity checks of symmetries are noiseless). In \Cref{sec:numerics} we numerically investigate the fidelity improvements from symmetry verification with noisy parity checks.

\subsection{Errors that commute with checks cannot be detected}
When performing a parity check with a symmetry $S$, not every error $E$ can be detected. If an error $E$ commutes with  $S$, then a check using $S$ cannot detect it. The reason is that the action of $E$ on the noise-free state $|\psi\rangle \langle \psi |$ keeps it within the symmetry-stabilized subspace: $P_{+1} E \ket{\psi} = E P_{+1} \ket{\psi} = E \ket{\psi}$. Therefore, to quantify the performance of SV with a symmetry $S$, we need to count the errors that commute with the symmetry. 

To this end we use a convenient binary symplectic notation~\cite{nielsen2011quantum}. In this notation an arbitrary tensor product of Pauli operators and identity operators $O$ is written in terms of length $N$ binary vectors $\vec{\alpha},\vec{\beta}$:
\begin{equation*}
O = O(\alpha,\beta) = i^{\alpha.\beta}\left(  X^{\alpha_1} \otimes \cdots \otimes X^{\alpha_N} \right) \left(  Z^{\beta_1}\otimes...\otimes Z^{\beta_N} \right).
\end{equation*}
In particular, the symmetry $\SB = \prod_{i} X_i$ is $\alpha=\vec{1},\beta=\vec{0}$. 

The advantage of this notation is that the commutation relation can be written as
\begin{align*}
[O(\alpha,\beta),O'(\alpha',\beta')] & = 0
\end{align*}
if and only if
\begin{align*}
\alpha.\beta' - \alpha'.\beta & \equiv 0 \mod{2}.
\end{align*}
The set of errors $O(\alpha,\beta)$ that commute with $\SB$ can now be described by the set of $\alpha,\beta$ that satisfy
\begin{equation}\label{undetectable Bit-Flip Binary notation}
\alpha.\vec{0}-\vec{1}.\beta = -\vec{1}.\beta \equiv 0 \mod{2}.
\end{equation}
The solution to this equation is the set of $\beta$  with an even number of $1$ as entries and $\alpha$ being an arbitrary binary vector.

\subsection{Warm-up: one QAOA layer with one local error}

We first consider the case where only a single error occurs after a single layer of QAOA. This highlights our point about undetectable errors and  allows us to understand the relative advantage gained by verifying $\SB$ and $S_{i,j}$. Let $E_a$ be the error acting on $a^{th}$ qubit. Because $\ket{\psi}$ is stabilized by $\SB$ and $\SB$ and $E_a$ either commute or anticommute, the following identities hold:
\begin{align}
\bra{\psi} E_a^\dagger \SB E_a \ket{\psi}  = & \begin{dcases}
1 & \text{for} \quad [E_a, \SB ] = 0, \\
-1 & \text{for} \quad \{ E_a, \SB \} = 0.
\end{dcases}
\end{align}
The ratio of fidelities \eqref{ratio fidelity} is straightforward to calculate for the depolarizing error channel \eqref{noisy channel 1 qubit}: 
\begin{equation}\label{eq:bitflip}\small
\frac{F_{\SB}}{ F_\text{noisy} }= \frac{1}{\Tr \left( P_{+1} \mathcal{E}_i(\rho) \right)} = \frac{1}{\sum_a \bra{\psi} E_a^\dagger \frac{\left(  I+S \right)}{2} E_a\ket{\psi}} = \frac{1}{1-\frac{2p}{3}}.
\end{equation}
Let us compare this with symmetry verification by a single permutation on the two qubits $i,j$. We must look at the set of errors $E_a$ that commute with the symmetry $S_{i,j}$.
Unlike $\SB$, $S_{i,j}$ is not in the Pauli group, so it is no longer necessarily true that it only commutes or anticommutes with the errors under  consideration. However, for the cases $E_a=X_i,Y_i,Z_i$, respectively, we observe the following:
\begin{equation*}\small
\bra{\psi} E_a S_{i,j} E_a \ket{\psi}  =  \begin{dcases}
\frac{1}{2}\bra{\psi}  \left( I_i I_j + X_i X_j-Y_i Y_j-Z_i Z_j \right) \ket{\psi} & \\
\frac{1}{2}\bra{\psi}  \left( I_i I_j - X_i X_j + Y_i Y_j-Z_i Z_j \right) \ket{\psi} & \\
\frac{1}{2}\bra{\psi}  \left( I_i I_j-X_i X_j-Y_i Y_j+Z_i Z_j \right) \ket{\psi} &
\end{dcases}
\end{equation*}
Notice the following useful identity:
\begin{align*}\small
\bra{\psi} X_i S_{i,j} X_i \ket{\psi} + \bra{\psi} Y_i S_{i,j} Y_i \ket{\psi} +\bra{\psi} Z_i S_{i,j} Z_i \ket{\psi} =1.
\end{align*}
Assume that the depolarizing noise channel \eqref{noisy channel 1 qubit} acts on qubit $i$, followed by SV with the symmetry $S_{i,j}$. Then we can write
\begin{equation}\label{eq:swap}\small
\frac{F_{S_{i,j}}}{ F_\text{noisy} }= \frac{1}{\Tr \left( P_{+1} \mathcal{E}_i(\rho) \right)} = \frac{1}{\sum_a \bra{\psi} E_a^\dagger \frac{\left(  I+S \right)}{2} E_a\ket{\psi}} = \frac{1}{1-\frac{p}{3}}.
\end{equation}
Comparing \eqref{eq:bitflip} with \eqref{eq:swap} suggests that the bit-flip parity check is more effective than the permutation parity check on two qubits in detecting depolarizing errors on a single qubit. This provides an intuition for why we expect permutation symmetries to be less powerful in mitigating errors, an effect that has been observed numerically in simulation and on superconducting qubit hardware~\cite{shaydulin2021error}.  %

\subsection{One QAOA layer with the full error model}
We now analyze the improvement in fidelity from bit-flip symmetry verification under local noise acting on all qubits. We begin with the single-layer case given by \eqref{full error model 1 layer} and then generalize to an arbitrary number of layers $d$. 
The goal is to understand the scaling behavior of the fidelity of the postselected state with number of qubits $N$ and depth $d$, for a given probability of error $p$. 

To calculate the improvement in fidelity from verifying a given symmetry \eqref{ratio fidelity}, we need to find the expectation $\Tr \left( P_{+1} \mathcal{E}\left( \rho \right) \right)$.
Only undetectable errors (those that commute with $\SB$) contribute.
We begin with the cases of depolarizing noise and dephasing noise for depth $d=1$.

\subsubsection{Depolarizing noise}
Let $K_{j_i}^{N_i}$ denote the Kraus operator labeled by $j_i$, which acts as the identity operator on all qubits except $N_i$. In particular, for depolarizing noise, the operators $K_{1}^{N_i},K_{2}^{N_i},K_{3}^{N_i}$ are $X_{N_i},Y_{N_i},Z_{N_i}$. respectively.

We follow the discussion in \cite{Marshall_2020} and label the noisy state by the number of error operators $m$ that have acted on it:
\begin{equation}
\ket{\psi}_\text{noisy}^m= \sum_{\vec{N}_m,\vec{j}_m} \psi_{\vec{j}_m}^{\vec{N}_m}= \sum_{\vec{N}_m,\vec{j}_m}K_{j_1}^{N_1}...K_{j_m}^{N_m} \ket{\psi}\,,
\end{equation}
where $\vec{N}_m,\vec{j}_m$ are length $m$ vectors.
Up to $N$ single-qubit errors can occur. To find the action of the noise channel, we sum the probabilities of errors occurring on $m$ qubits at a time:
\begin{equation}\label{analytic error model depolarizing}
   \mathcal{E}(\rho)= \sum_{m=0}^{N} (1-p)^{N-m} \left( \frac{p}{3} \right)^m \ket{\psi}_\text{noisy}^m\bra{\psi}_\text{noisy}^m.
\end{equation}
Inserting the projector onto the $+1$ eigenspace yields
\small
\begin{equation*}
\Tr \left( P_{+1} \mathcal{E}(\rho) \right) = \sum_{m=0}^{N} (1-p)^{N-m} \left( \frac{p}{3} \right)^m \bra{\psi}_{\text{noisy}}^m \frac{1}{2}\left(I+S \right)\ket{\psi}_{\text{noisy}}^m.
\end{equation*}
\normalsize
Whenever $[\SB,K_{j_1}^{N_1} \cdots K_{j_m}^{N_m}] = 0$, the contribution from matrix element $\bra{\psi}_{\text{noisy}}^m \frac{1}{2}\left(I+S \right)\ket{\psi}_{\text{noisy}}^m$ is one; and when they anticommute, it is zero.

To count the nonzero contributions, we rewrite the sum as
\begin{equation}
\sum_{m=0}^{N} (1-p)^{N-m} \left( \frac{p}{3} \right)^m f(m),
\end{equation}
where $f(m)$ counts the number of operators that commute with $\SB$. In our analysis of undetectable errors \eqref{undetectable Bit-Flip Binary notation}, these operators are described by pairs of length $N$ binary vectors $\left( \alpha, \beta \right)$ such that $\alpha$ is arbitrary and $\beta.\vec{1} \equiv 0 \mod{2}$. Counting the number of such vectors $f(m)$ is a problem that we solve in Appendix~\ref{depolarizing appendix}. The result is
\begin{align*} 
  f(m)&= \frac{1}{2} \binom{N}{m} \left( 1 + 3^m \right) \text{ for }m\text{ even.} \\ \nonumber
  &= \frac{1}{2} \binom{N}{m} \left( -1 + 3^m \right) \text{ for }m\text{ odd,}%
\end{align*} 
 $f(m)$ yields the ratio of fidelities for depolarizing noise at a depth $d=1$:
\begin{equation}\label{depolarizing error imp}
\frac{F_{\SB} }{F_\text{noisy}} = \frac{1}{\sum_{m=0}^{N} (1-p)^{N-m} \left( \frac{p}{3} \right)^m f(m)}=\frac{1}{\mathcal{F}(N,p)},
\end{equation}
where
\begin{widetext}
\begin{align*}
\mathcal{F}(N,p)=& \frac{1}{4} \left( 1 + (1 - 2 p)^{N} + (1 - \frac{4 p}{3})^{N } + (1 - \frac{2 p}{3})^{N }\right)  + \frac{1}{4} (1 - p)^{1 + N } (-1 + p)^{-1 - N } \\ \nonumber
&\left(-1 + (1 - 2 p)^{N } + (-1 + p)^{N }(1 - \frac{p}{(3 (-1 + p))})^{N }) - (-1 + p)^{N } (1 + \frac{p}{(3 (-1 + p))})^{N }\right).
\end{align*}
\end{widetext}
\begin{figure}
\includegraphics[width=\linewidth]{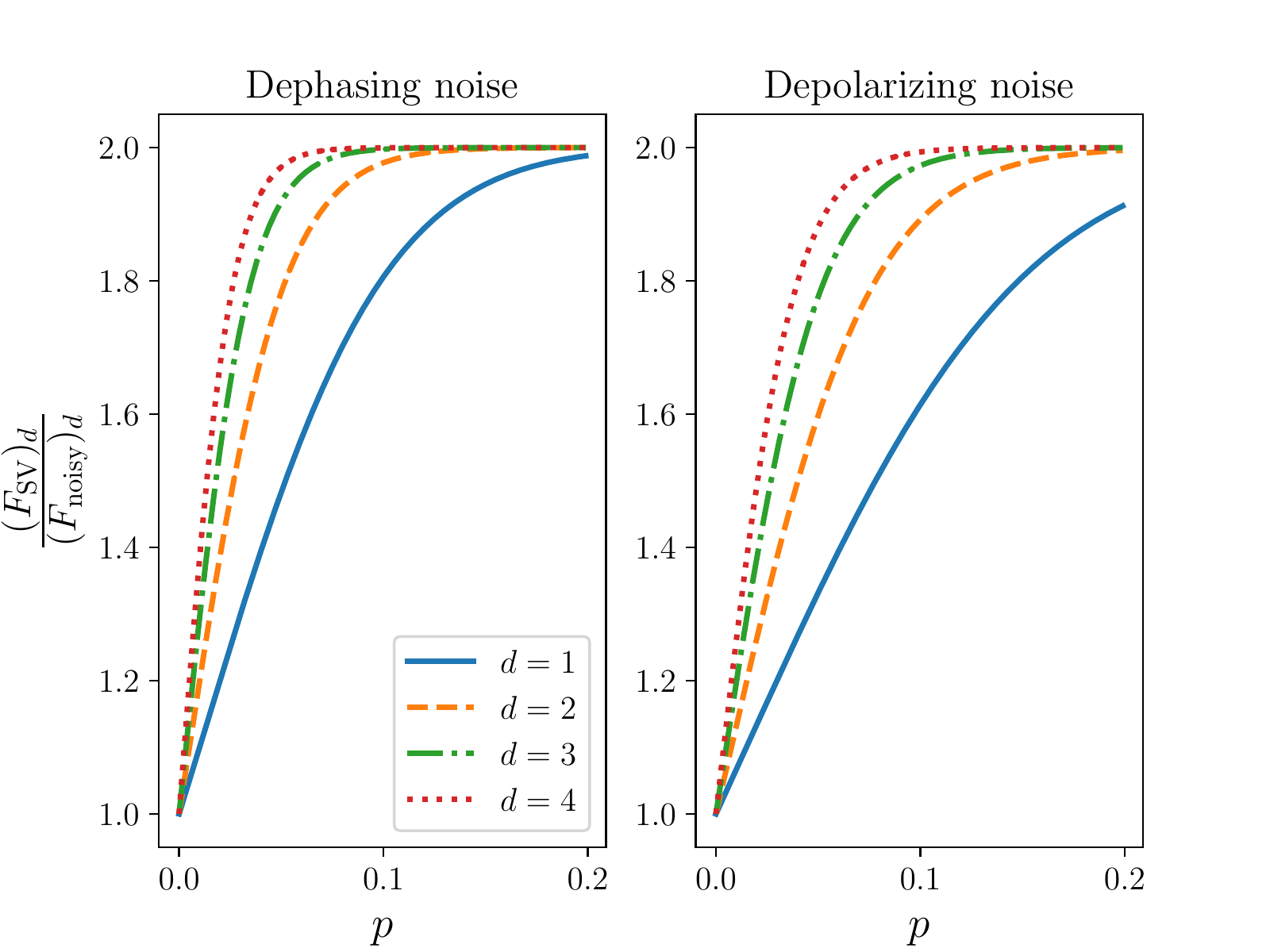}
\caption{Ratio of fidelities with  and without SV for fixed $N=10$ for varying depth. The value on the $y$ axis is produced by \Cref{eq: dephasing plotted,eq: depolarizing plotted}, which are based on the local noise model depicted in \Cref{fig:circuit2}.}\label{fig:analytic}
\end{figure}

\subsubsection{Dephasing channel noise}
We now consider the case of depth $d=1$ with the dephasing noise channel described by \eqref{noisy channel 1 qubit dephasing} acting on each qubit.
Similar to the previous case, the expectation value takes the form
\begin{equation}
\Tr \left( P_{+1} \mathcal{E}(\rho) \right) = \sum_{m=0}^{N} (1-p)^{N-m} \left( p \right)^m f(m).
\end{equation}
The crucial difference is in the nonzero contribution to $f(m)$ in the evaluation of  $[\SB,K_{j_1}^{N_1}\cdots K_{j_m}^{N_m}]$. The errors are purely $Z$ errors at multiple sites. In the binary symplectic notation, tensor products of $Z$ operators satisfy $\vec{\alpha} = \vec{0}$. %
Thus, the only contribution to the sum arises from errors %
with $\vec{\alpha} = \vec{0}$ and $\vec{\beta}$ satisfying $\vec{\beta}.\vec{1} \equiv 0 \mod{2}$. The number of such errors is given by $f(m)=  \binom{N}{m}$, where $m$ is constrained to be even. The ratio can be computed as follows:
\begin{equation}\label{phase fidelity imp}
\frac{F_{\SB}}{F_\text{noisy}} = \frac{1}{\displaystyle \sum_{m \text{  even}}^{N} (1-p)^{N-m} p ^m f(m)}= \frac{2}{(1 + (1 - 2 p)^N)}.
\end{equation}

\subsubsection{Arbitrary depth $d$}
We now extend our results to a higher depth $d$ under the full noise channel described by \eqref{full noise model}. Similar to the simpler $d=1$ case, the errors that are not detected commute with $\SB$. 
One still needs only to compute the ratio 
\begin{equation}
\frac{\left( F_{\SB} \right)_d}{\left( F_\text{noisy}\right)_d}=\frac{1}{\Tr \left( P_{+1} \rho_d \right)}.
\end{equation}
The noisy state $\ket{\psi}_\text{noisy}^m$ can be described by
\begin{equation*}
\small
\ket{\psi}_\text{noisy}^m= \sum_{\vec{N}_m,\vec{j}_m} \psi_{\vec{j}_m}^{\vec{N}_m}= \sum_{\vec{N}_m,\vec{j}_m}K_{j_1}^{N_1}\cdots K_{j_m}^{N_m}\ket{\psi},
\end{equation*}
where the length of the vector $\vec{N}_m$ can be at most $Nd$ instead of $N$ because the error could have occurred at $Nd$ locations. 
Because we assume that the noise channel factorizes over qubits and the channel acting on each qubit and depth is identical, $\rho_d$ is effectively calculated by replacing $N \rightarrow Nd$ in the result, as is done in \cite[Eq.~(16)]{Marshall_2020}. 
The end result is

\small
\begin{equation}\label{eq: dephasing plotted}
\frac{\left( F_{\SB} \right)_d}{\left( F_\text{noisy}\right)_d} = \frac{1}{\displaystyle \sum_{m \text{ even}}^{Nd} (1-p)^{Nd-m} p ^m f(m)} \approx \frac{2}{(1 + (1 - 2 p)^{Nd})}.
\end{equation}
\normalsize
For the depolarizing noise channel, a similar analysis gives
\begin{equation}\label{eq: depolarizing plotted}
\frac{\left( F_{\SB} \right)_d}{\left( F_\text{noisy}\right)_d} \approx \frac{1}{\mathcal{F}(Nd,p)},
\end{equation}

\subsubsection{Qualitative observations}
The ratios of fidelities for depolarizing and dephasing error channels are plotted in \Cref{fig:analytic} and have interesting features. First, for a fixed error rate $p$ and fixed $N$, the fidelity boost provided by SV increases with the depth $d$. This increase is because there is a larger probability for a number of detectable errors to have accumulated in the circuit. 

A second important feature is that for fixed values of $p$ and $d$, $\frac{\left(F_{\SB}\right)_{d} }{\left(F_{\text{noisy}}\right)_d}$ saturates to $2$ as $N \rightarrow \infty$. This suggests that in its current form, error mitigation with SV is not scalable to large $N$ because the cost of applying the parity check via $N$ $\cnotgate$ gates scales linearly with $N$, while the fidelity boost saturates.

\section{Verifying higher-order permutation symmetries}\label{sec:higher order permutations}

In \Cref{sec:background} problem-instance-specific permutation the symmetries were defined. We now address the question of whether verifying these symmetries for a given problem instance can give a significant fidelity boost.

To address this question, we must first calculate the projection operator onto the symmetry-stabilized subspace. Every permutation of a finite set can be written as a cyclical permutation (cycle) or as a product of disjoint cycles. Constructing a set of orthogonal and complete projectors for each cycle is straightforward because each cycle $s$ of length $q$ must satisfy $s^q=1$. The operator representing $s$ on qubits satisfies $S^q = I$. As a result, the eigenvalues of this operator lie on the unit circle and are given by $e^{2 \pi i \frac{k}{q} }$, where $k\in \{0,...,q-1\}$. The symmetry-stabilized subspace corresponds to the $k=0$ case, which is the subspace corresponding to the $+1$ eigenvalue. 

To find the projector onto the $+1$ eigenspace of $S$, we factorize the following equation: 
\begin{equation}
S^q - I = \left( \Phi_q(S) \right) \left( S - I \right)=0.
\end{equation}
$\Phi_q(x)$ is the cyclotomic polynomial, which is the unique divisor of the polynomial equation $x^q-1$. These polynomials are known in closed form. The final result is that $\Phi_q(S)$ projects vectors onto the $+1$ eigenspace of $S$. 

To illustrate this point, we consider the permutation on three elements, denoted by the cycle $(1,2,3)$. This satisfies $s^3-1=(1+s+s^2)(s-1)=0$. The associated operator $S$ can be implemented by successive $\swapgate$s $S=\swapgate_{1,2}. \swapgate_{2,3}$, and the projector onto the $+1$ eigenspace is 
\begin{equation}\label{eq:projector_permutation_ex}
P_{+1} = \frac{1}{3} \left( I+ S + S^2 \right).
\end{equation}
To see that \eqref{eq:projector_permutation_ex} is correct, consider applying $P_{+1}$ on any state $\ket{v}$ in the subspace associated with the eigenvalues $e^{2 \pi i \frac{1}{3}},e^{2 \pi i \frac{2}{3}}$. Then $P_{+1} \ket{v}=0$. This is because $(1+s+s^2)=0$ for $s=e^{2 \pi i \frac{1}{3}}\ , e^{2 \pi i \frac{2}{3}}$. 
The normalization is chosen to ensure that the corresponding eigenvalue is $1$. 

To construct the projector circuit, we define the shifted operator
\begin{equation}
U = \frac{1}{3} \left( 2S+2S^2 - I \right).
\end{equation}
This operator is defined so that applying controlled-$U$ and then  postselecting the $+1$ eigenvalue projects onto the symmetry-stabilized subspace. Operator $U$ is not trivial to decompose into gates in general. For simple cases like the example considered above, circuit synthesis tools such as QSearch \cite{Qsearch} can be used. 

For general cycles, however,  the corresponding projector is difficult to decompose into gates. In this case, $\left( \Phi_q(S) \right)$ gives an expression for the projector as a linear sum of permutation operators. One may leverage the techniques for implementing linear combinations of unitaries outlined in \cite{childs2012addingunitaries}.  %

In general, for a given symmetry this task of projection onto the $+1$ eigenspace can be done by performing quantum phase estimation (QPE) and postselecting on the measurement outcome $0^{N}$, which corresponds to measuring out eigenvalue $+1$. The resulting postselected state would be the state projected into the $+1$ eigenspace. Of course, using QPE for error mitigation is impractical.  The high gate cost of QPE is due to the need to apply $S$, $S^2$, \ldots, $S^{2^k}$, controlled on the ancilla qubit, where $k$ is the number of binary digits sufficient to distinguish the $+1$ eigenvalue with some target probability.

Symmetry verification of higher-order permutations may be beneficial, but without further advances it may not be viable because of the high overhead of performing the symmetry checks. One promising direction is generalizing the efficient postprocessing methods that do not rely on directly verifying the symmetries using projectors, such as those developed in \cite{McClean} for symmetries with generators in the Pauli group.

\section{Experiments in Simulation}\label{sec:numerics}
In this section we present the results from numerical simulations of symmetry verification. We assign error rates to all single- and two-qubit gates, including those used to perform the parity checks. 
The circuit is first decomposed into a set of native single-qubit gates $R_x$, $R_y$, and $R_z$ and the two-qubit gate $\cnotgate$. The following error channel acts every time a gate is applied, with $k=1,2$ for single- and two-qubit gates, respectively:
\begin{equation}\label{qiskit error model}
    \mathcal{E} \big( \rho \big) = \left( 1-p_k\right) \rho +p_k\frac{\Tr\left( \rho \right)}{2^n} I.
\end{equation}

The error rates associated with single-qubit gate fidelities are typically smaller than those of two-qubit gate fidelities by an order of magnitude on common hardware platforms~\cite{ibm_qc_website,ionq_website}. To model this, we set $p_1=\frac{1}{10}p_2$. %

We use preoptimized QAOA parameters from noiseless simulations because the noise model we use does not affect the value of optimal QAOA parameters \cite{xue2019effects,2021NIBP}. For depth $d\leq 3$ we use parameters available in QAOAKit \cite{shaydulin2021qaoakit}, and for $d > 3$ we find them by using exhaustive numerical optimization. 

\begin{figure*}
\includegraphics[width=0.91\linewidth]{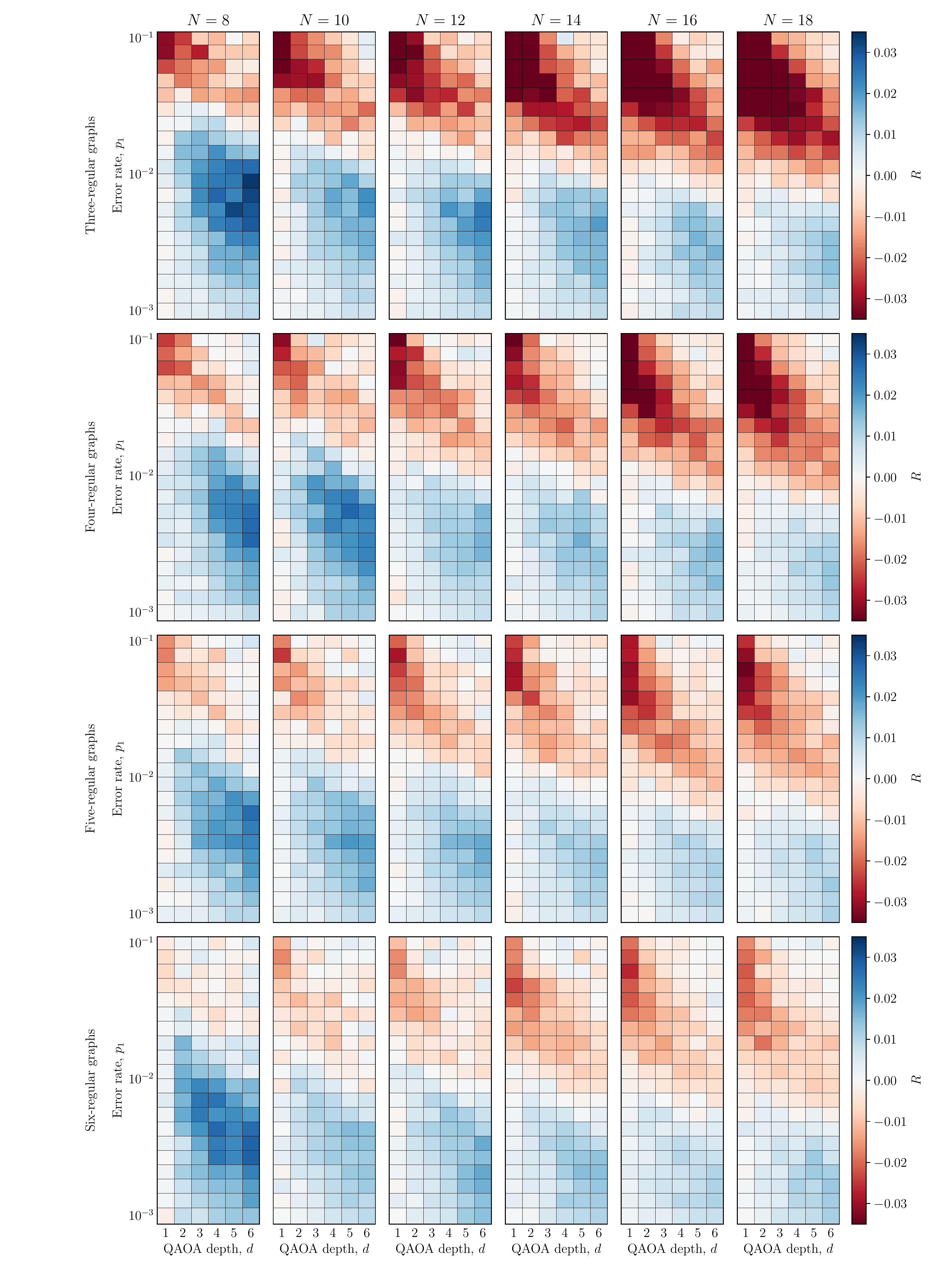}
  \caption{Improvement in approximation ratio $R$ given by \Cref{eq: definition of R} for varying number of nodes $N$, QAOA depth $d$, and error rate $p_1$. We see a greater improvement in $R$ from the application of SV for larger QAOA depth. For larger qubit count $N$, lower single-qubit error rates are required in order to see an improvement in QAOA objective.}
  \label{fig:all_simulation}
\end{figure*}

To quantify the effect of errors on the QAOA circuit, we consider the expectation value of the cost function in the noisy state $\langle H \rangle$ given by \eqref{eq:obj}. Our results are presented in \Cref{fig:all_simulation}.
We consider random regular graphs with number of nodes $N \in \{ 8,10, \ldots, 18\}$, degree $\in \{3,4,5,6\}$, and  depth $d\in \{1,2,3,4,5,6\}$. To map out the space of error rates where SV shows an improvement, we consider the following quality measure of improvement after symmetry verification:
\begin{equation}\label{eq: definition of R}
R=\frac{\langle H \rangle_{\text{SV}}}{\langle H \rangle_{\text{no-SV}}}-1.
\end{equation}
When $R>0$, we see an improvement in fidelity by applying SV (indicated by the color blue in \Cref{fig:all_simulation}). 

For each degree and number of nodes, we observe a clear regime of error rates and circuit depths in which symmetry verification leads to improvements in the QAOA objective. We see that this regime corresponds to realistic error rates that can be expected on near-term hardware. Concretely, \Cref{fig:all_simulation} shows that when $\cnotgate$ error rates are (realistically) between 1 and 10\%, using SV is beneficial. The improvement from applying SV increases with QAOA depth, which is consistent with our analysis in \Cref{sec:analytics}. Moreover, for higher qubit counts, the error rates required for obtaining improvement from SV are lower. The reason is  the cost of applying $N$ $\cnotgate$ gates during the parity check, which grows with the number of nodes.

\section{Experiments on IonQ hardware for small $N$}\label{sec:ionQ}
To verify the results obtained in simulation, we tested error mitigation using our symmetry verification approach on an 11-qubit IonQ trapped-ion quantum computer \cite{2019ionq} with an all-to-all connectivity. Each qubit register consists of a chain of spatially confined 171Yb+ ions where the quantum states are encoded into the hyperfine sublevels of the atom. 

\begin{figure*}
\centering
  \includegraphics[width=\linewidth]{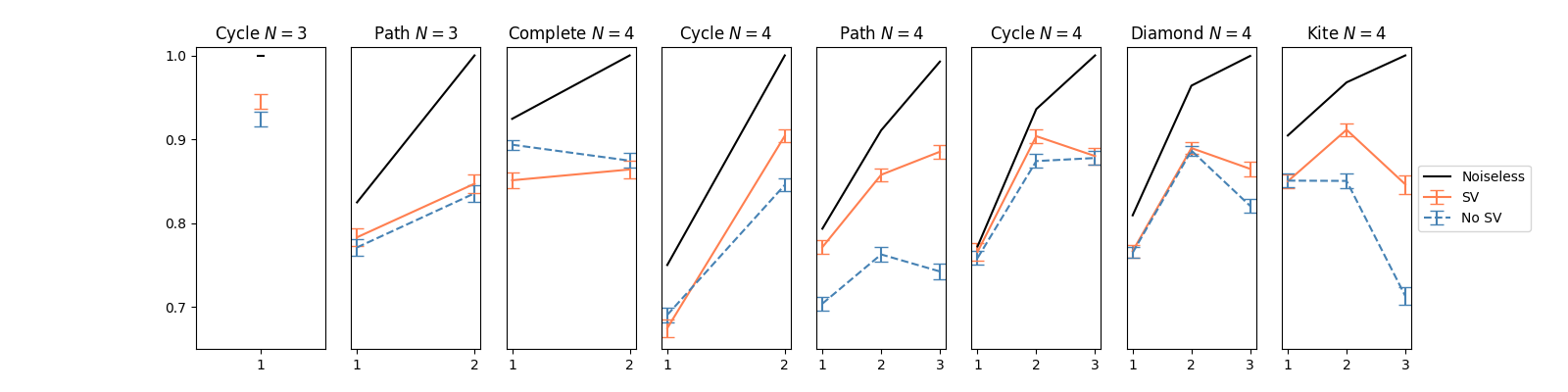}\\
  \hspace{-0.5in}\includegraphics[width=0.1\linewidth]{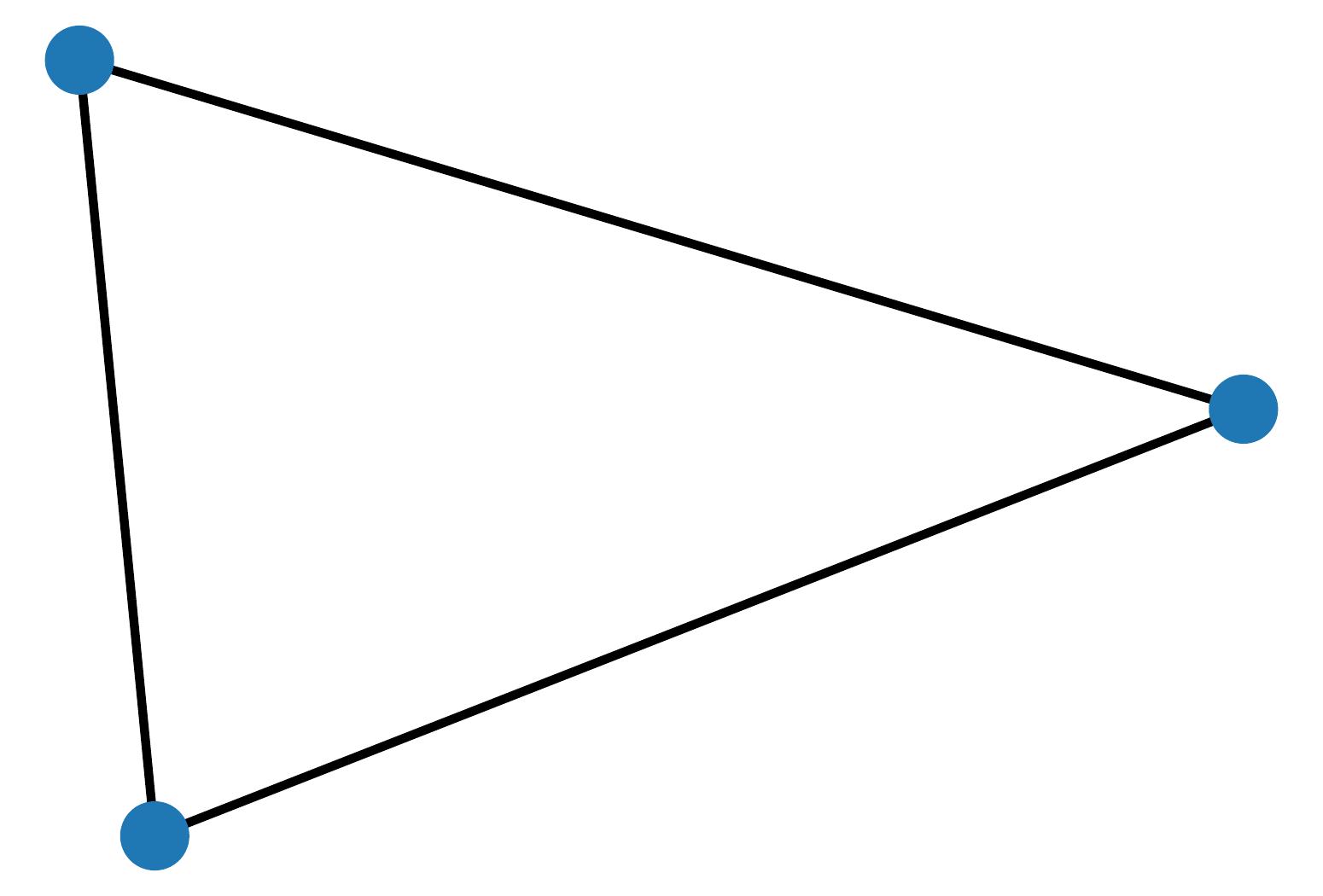}
  \includegraphics[width=0.1\linewidth]{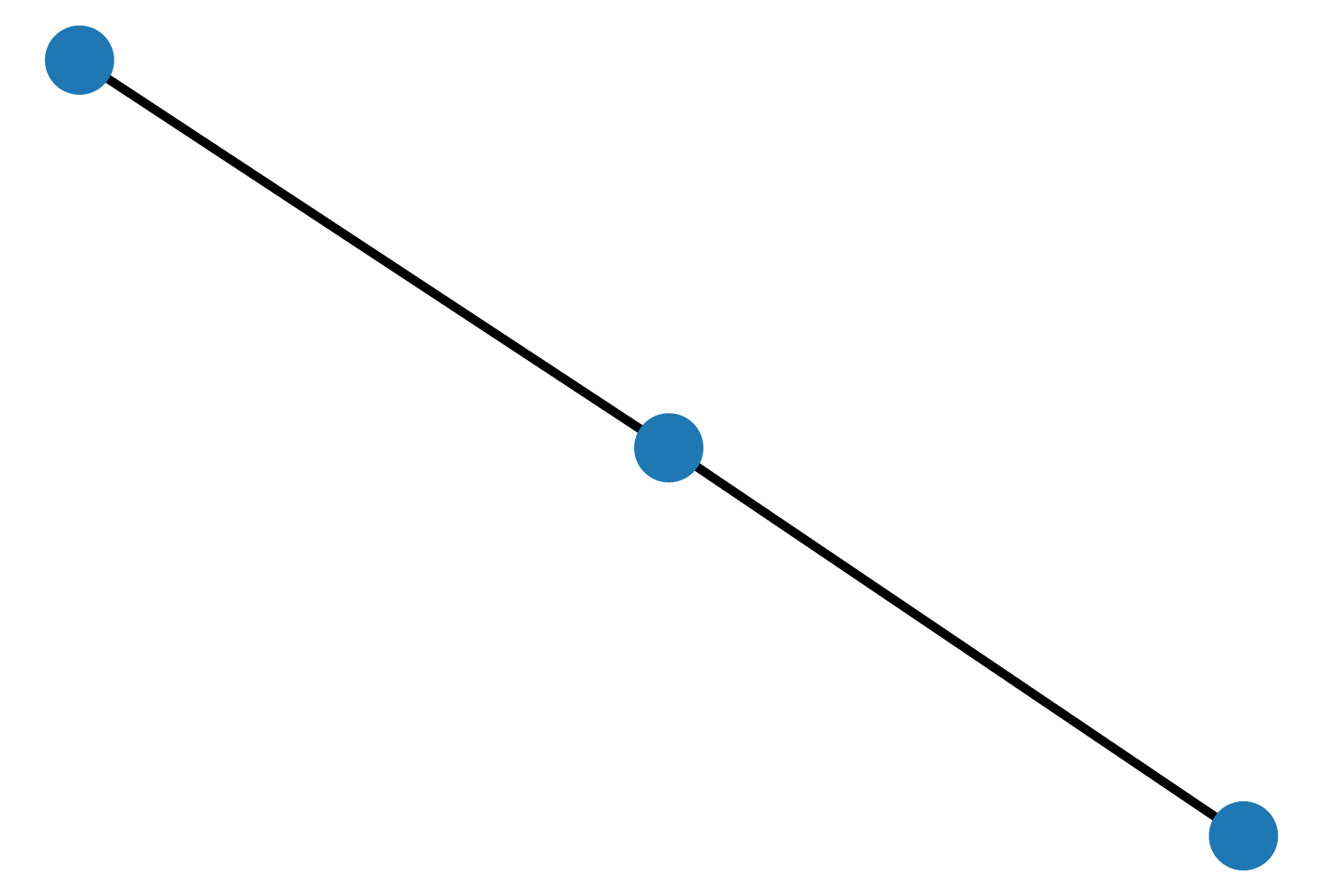}
  \includegraphics[width=0.1\linewidth]{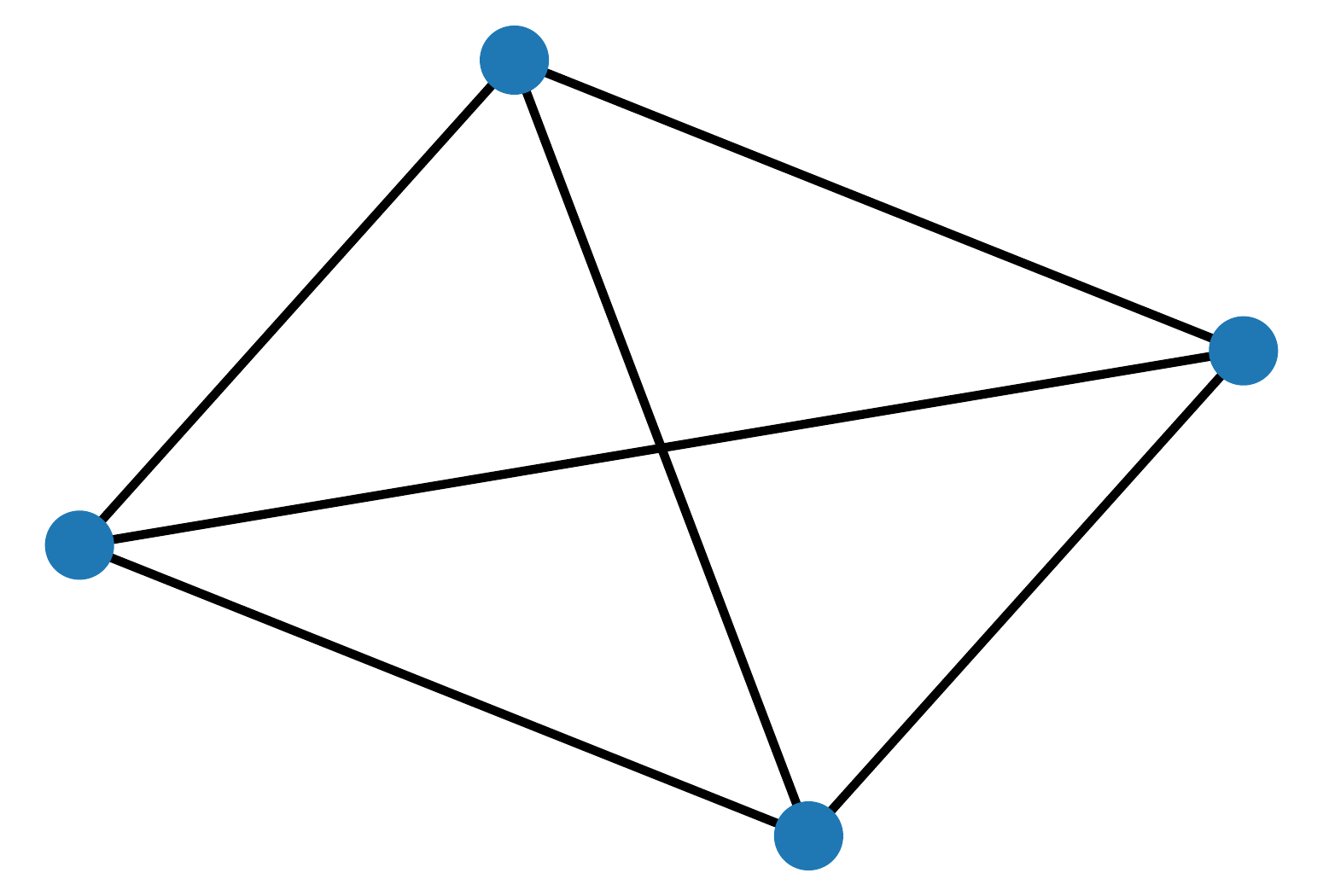}
  \includegraphics[width=0.1\linewidth]{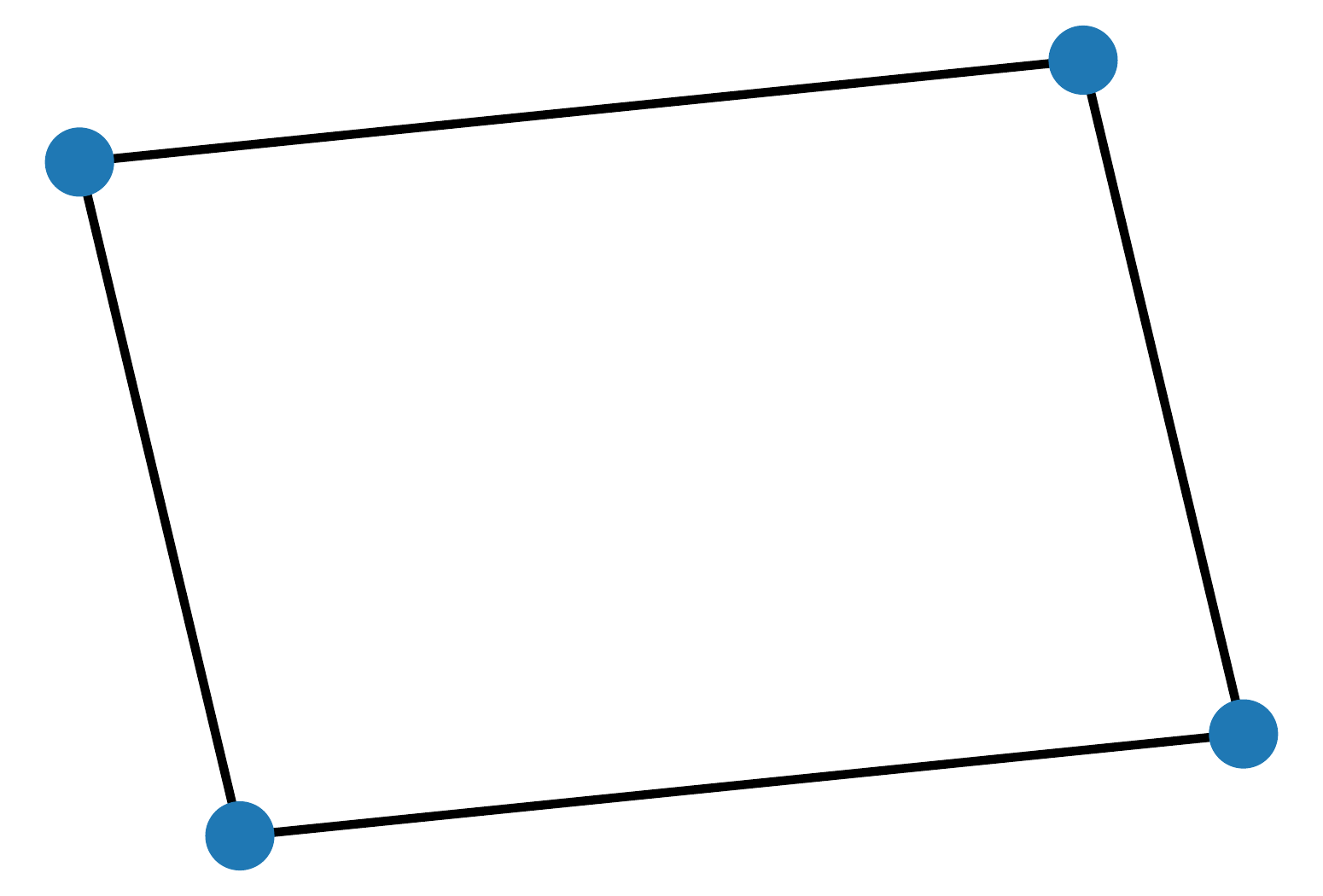}
  \includegraphics[width=0.1\linewidth]{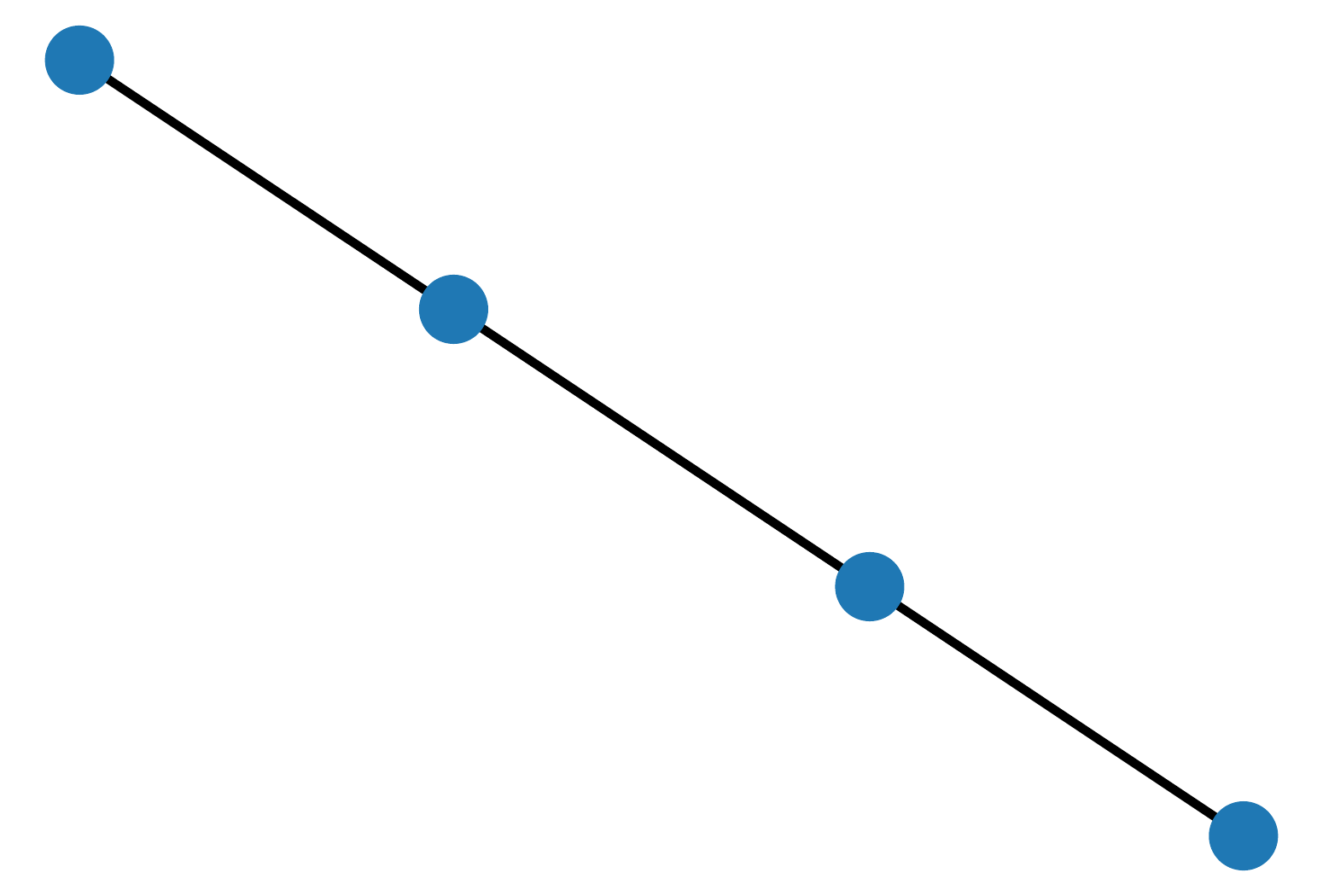}
  \includegraphics[width=0.1\linewidth]{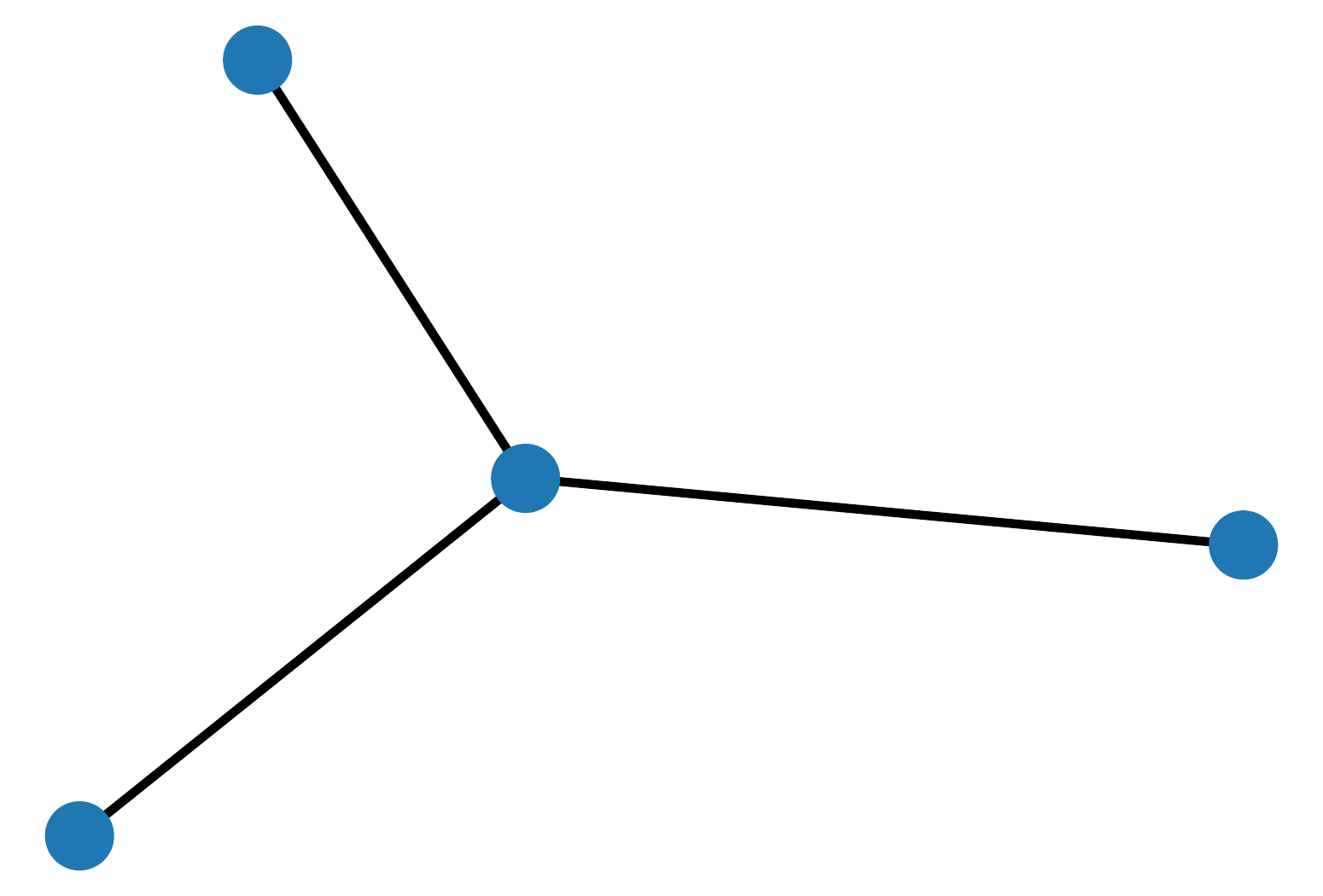}
  \includegraphics[width=0.1\linewidth]{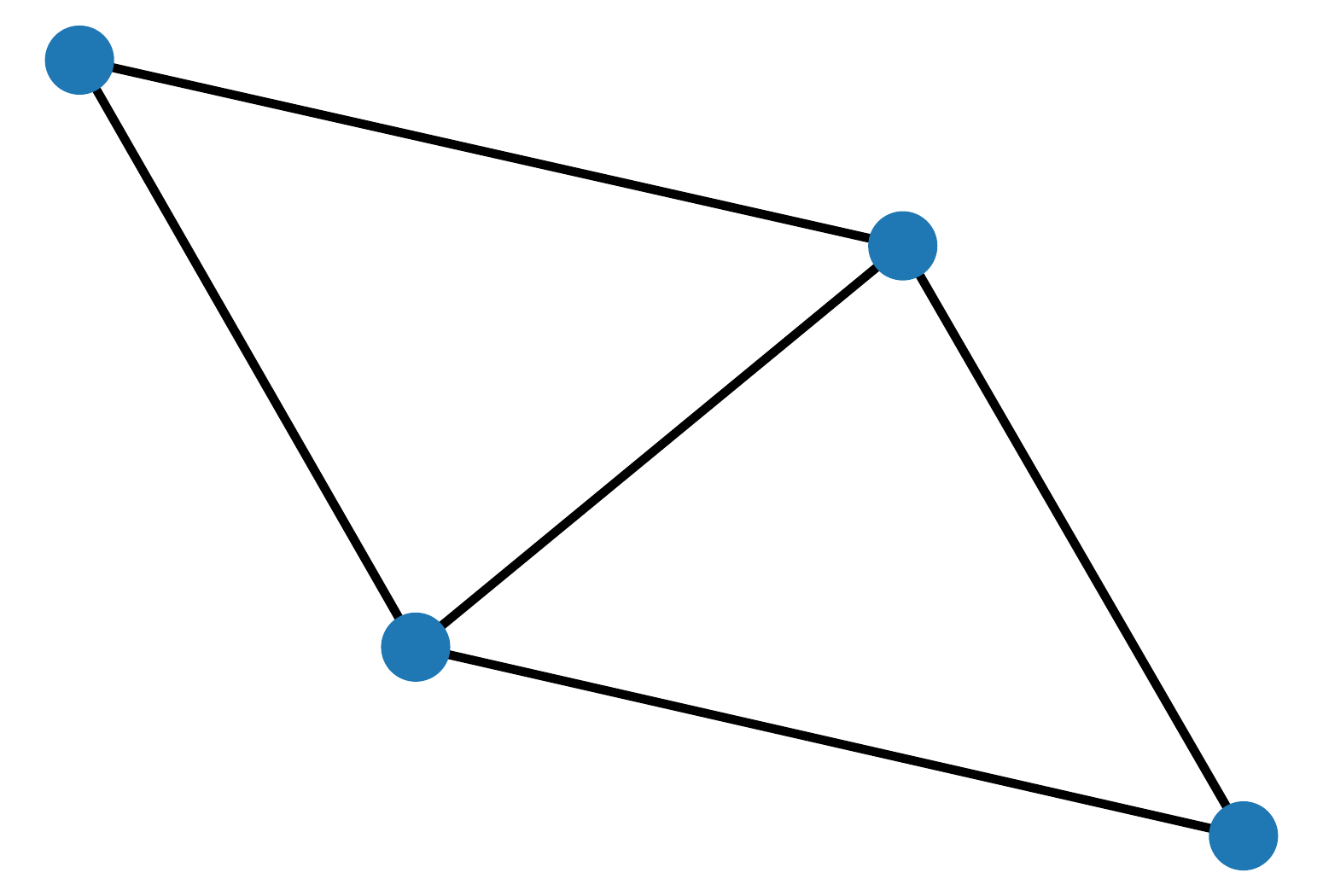}
  \includegraphics[width=0.1\linewidth]{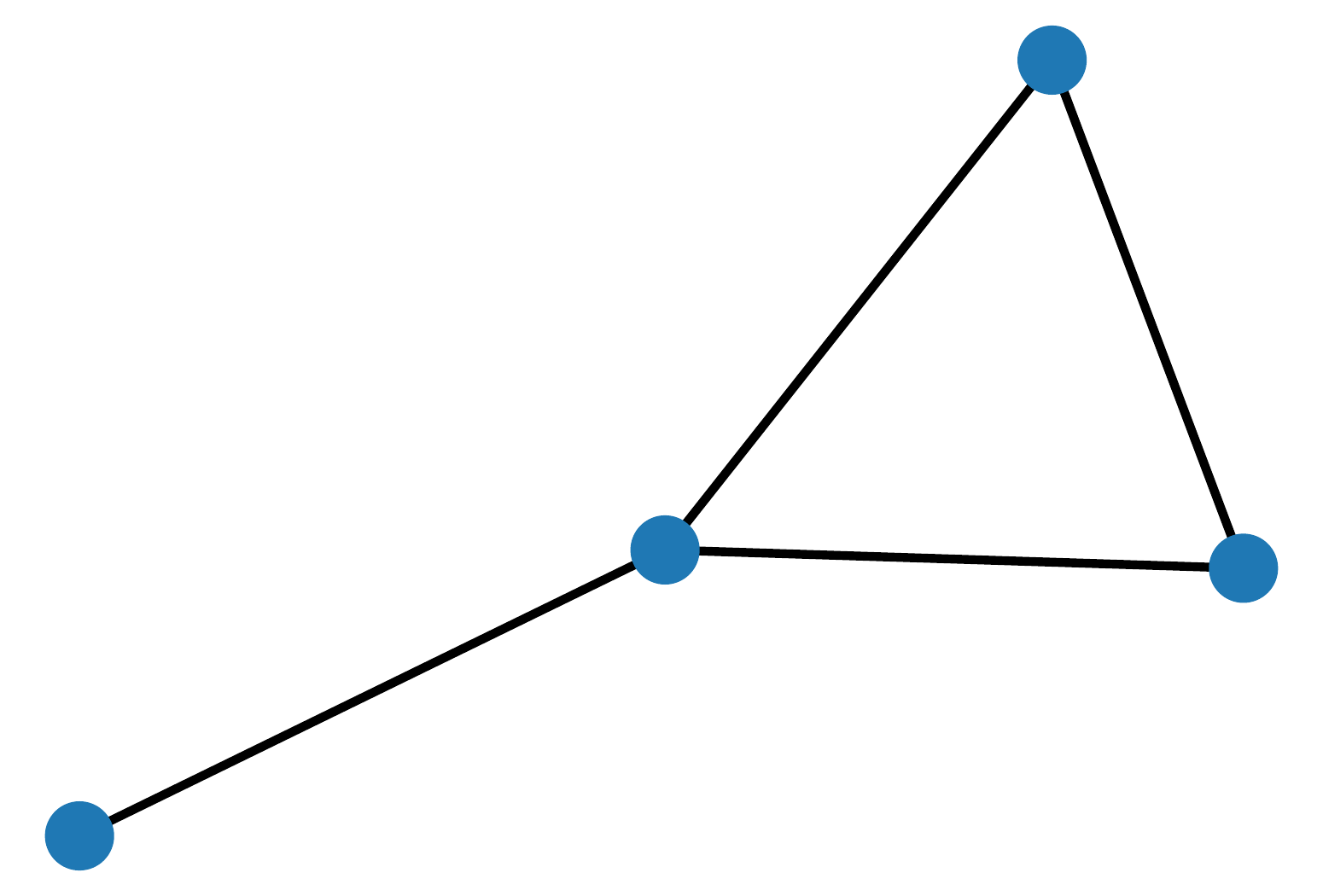}
  \caption{Results of experiments done on trapped-ion quantum hardware made accessible by IonQ. We plot the approximation ratio after applying SV to QAOA applied to MaxCut (shown in orange), without applying SV (shown in blue), and noiseless simulation (shown in black). We observe an improvement after applying SV in all instances except Complete $N=4$.}
  \label{fig:all_ionq}
\end{figure*}

We ran experiments using all nonisomorphic graphs on $3$ and $4$ nodes with the QAOA depth being  1, 2, or 3.
The results are shown in \Cref{fig:all_ionq}.
We observe improvement in the QAOA objective for most graphs and depths. The change in the approximation ratio from SV varies from a decrease of $4.5\%$ to an improvement of $19.2\%$, with a mean improvement of $4.3\%$. The results observed on hardware align well with numerical simulations, since typical two-qubit gate error rates reported for IonQ trapped-ion systems are between $1\%$ and $3\%$, matching the error rates for which our simulation experiments predicted improvement from SV.

We need to put into perspective the improvement in approximation ratio observed on IonQ hardware (up to $19.2\%$) and in numerical simulations with local noise (up to $3.4\%$). For  3-regular graphs, Farhi et al.~\cite{farhi2014quantum} showed that using the QAOA ansatz, the approximation ratio at depth $d=1$ must be no less than $0.692$, and Wurtz et al.~\cite{2021Wurtz} showed an improved bound of $0.7559$ for $d=2$. Assuming a conjecture on the graphs that are the worst case (with no “visible” cycles), Wurtz et al.~\cite{2021Wurtz} also extended this bound to $0.7924$ for $d=3$. The best-known classical algorithm by Goemans and Williamson \cite{Goemans1995ImprovedAA} lower bounds the approximation ratios by $0.8786$ \cite{Frieze1995ImprovedAA}. At the same time, it is known that MaxCut is NP-hard to approximate better than $\frac{16}{17}\approx 0.941$ in the worst case~\cite{Hstad2001}. Therefore, a small (i.e., less than $10\%$) improvement in the approximation ratio can be essential to achieving quantum advantage.

\section{Discussion}\label{sec:discussion}
In the absence of full fault tolerance, error mitigation is required for the application of QAOA in practice. In this paper we demonstrate that symmetry verification applied to QAOA circuits improves the fidelity of the quantum state. We provide techniques for studying the scaling of error mitigation techniques with circuit size and depth. We use these techniques to demonstrate the scaling of symmetry verification for QAOA applied to MaxCut under local noise. Such scaling analysis is essential for the long-term goal of optimizing the use of error mitigation strategies. The methods we use in this work can be easily adapted to develop a similar understanding of error mitigation by symmetry verification in other classes of variational quantum algorithms, including the variational quantum eigensolver \cite{2014} and quantum alternating operator ansatz \cite{2021Streif}. For these algorithms, the symmetries include particle number conservation and local particle number conservation. 

Our analysis of local noise models suggests that verifying the bit-flip ($\mathbb{Z}_2$) symmetry of the cost function is useful when running QAOA circuits at higher depths. Because worst-case upper bounds on performance associated with QAOA become competitive with classical algorithms only at high depths, SV may become a useful part of the error mitigation stack of algorithms. We perform a set of numerical experiments  to simulate the effects of the infidelity of two-qubit gates.  These allow us to map out the parameter space where SV is expected to be useful. Our analysis via local error models indicates that higher-depth circuits show a greater increase in the fidelity of the final state. Experiments on IonQ hardware validate these findings.  We also observe that verifying this bit-flip symmetry is more efficient than verifying permutation symmetries of the cost function. This result is consistent with the experimental evidence from superconducting IBM quantum processors reported in \cite{Shaydulin2012}. 

As a part of our analysis, we made use of the binary symplectic representation commonly used to describe quantum error-correcting codes. An interesting future direction is to use this representation to appropriately modify the mixer Hamiltonian $B$ such that $B$ and $H$ are generators of a code subspace with a distance no less than 2. Then, one could perform parity checks that would detect all single-qubit errors and extend the reach of this analysis.

An important generalization to point out is Max-$k$-Cut, the problem of finding an approximate $k$-vertex coloring of a graph. Bravyi et al.~\cite{Bravyi2020HybridQA} studied the approximation ratio achieved by QAOA for Max-$k$-Cut and compared it with the best-known classical approximation algorithms. The global symmetries of this model are permutations $S_k$. We expect that verifying this larger symmetry after the circuit has been executed will allow us to detect  more errors and hence provide a greater boost in the fidelity. 

\section*{Acknowledgment}
This material is based upon work supported by the U.S. Department of Energy, Office of Science, National Quantum Information Science Research Centers, 
the Office of Advanced Scientific Computing Research, Accelerated Research for Quantum Computing program, and 
Laboratory Directed Research and Development (LDRD) funding from Argonne National Laboratory, provided by the Director, Office of Science, at the U.S. Department of Energy  
under contract number DE-AC02-06CH11357. We gratefully acknowledge the computing resources provided on Bebop, a high-performance computing cluster operated by the Laboratory Computing Resource Center at Argonne National Laboratory.

\bibliographystyle{IEEEtran}
\bibliography{cite.bib}

\appendices

\section{$f(m)$ for Depolarizing channel}\label{depolarizing appendix} 
Consider the case of even $m$. Because there are $m$ errors, we first consider errors with $\vec{\alpha}$ having Hamming weight of $i$ $\vec{\beta}$ having Hamming weight $m-i$. Note that $i$ must be even because of the condition $\vec{\beta}.\vec{1} \equiv 0 \mod{2}$. There are $\binom{N}{m-i}$ ways of picking the vector $\vec{\beta}$. There are $\binom{N-m+i}{i} 2^{m-i}$ ways of choosing the vector in $\vec{\alpha}$ because it must have an entry 1 in at least $N-(m-i)$ positions (for the total number of errors to be $m$), which gives a factor of $\binom{N-m+i}{i}$. The vector $\vec{\alpha}$ can have $0$ or $1$ in the remaining $m-i$ positions, which gives a factor of $2^{m-i}$. 
The full contribution to $f(m)$ takes the following form:
For even $m$, doing this sum gives
\begin{align*} 
  f(m) & = \sum_{i \text{ even}}^{m} \binom{N}{m-i} 2^{m-i} \binom{N-m+i}{i} \\
       & = \frac{1}{2} \binom{N}{m} \left( 1 + 3^m \right). %
\end{align*}
For odd $m$, the logic is the same as before except that we must consider errors for which $\vec{\beta}$ has a Hamming weight of $m-i$, where $i$ is odd to satisfy the constraint $\vec{\beta}.\vec{1} \equiv 0 \mod{2}$.
The contribution to $f(m)$ takes the following form:
\begin{align*} 
  f(m) & = \sum_{i \text{ odd}}^{m} \binom{N}{m-i} 2^{m-i} \binom{N-m+i}{i} \\
       & = \frac{1}{2} \binom{N}{m} \left( -1 + 3^m \right).
\end{align*}

\vfill
\framebox{\parbox{.90\linewidth}{\scriptsize The submitted manuscript has been created by
UChicago Argonne, LLC, Operator of Argonne National Laboratory (``Argonne'').
Argonne, a U.S.\ Department of Energy Office of Science laboratory, is operated
under Contract No.\ DE-AC02-06CH11357.  The U.S.\ Government retains for itself,
and others acting on its behalf, a paid-up nonexclusive, irrevocable worldwide
license in said article to reproduce, prepare derivative works, distribute
copies to the public, and perform publicly and display publicly, by or on
behalf of the Government.  The Department of Energy will provide public access
to these results of federally sponsored research in accordance with the DOE
Public Access Plan \url{http://energy.gov/downloads/doe-public-access-plan}.}}
\clearpage

\end{document}